**Here your Paper ID – 5 numbers**
Full Name of Study Committee –example A2 Power transformers and reactors
Preferential Subject - example PS1 Design of resilient transformers
*Information available from your National Committee and in the emails sent to your att.*
*delete before publication.*

## Scalable Neural Dynamic Equivalence for Power Systems

**Qing SHEN*[1], Yifan ZHOU[1], Huanfeng ZHAO[1], Peng ZHANG[1], Qiang Zhang[2], Slava Maslennikov[2], Xiaochuan Luo[2]**
1. Stony Brook University   2. ISO New England
Qing.shen@stonybrook.edu

**SUMMARY**

Traditional grid analytics are model-based, relying strongly on accurate models of power systems, especially the dynamic models of generators, controllers, loads and other dynamic components. However, acquiring thorough power system models can be impractical in real operation due to inaccessible system parameters and the privacy of consumers, which necessitate data-driven dynamic equivalencing of unknown subsystems. Learning reliable dynamic equivalent models for the external systems from SCADA and PMU data, however, is a long-standing intractable problem in power system analysis due to complicated nonlinearity and unforeseeable dynamic modes of power systems. This necessitates dynamic equivalencing for unknown subsystems, which employs physics-informed machine learning and neural ordinary differential equations (ODE-NET) to preserve dynamic behaviors post-disturbances. The new contributions are threefold:

- A NeuDyE formulation to enable a continuous-time, data-driven dynamic equivalence of power systems, saving the effort and expense of acquiring inaccessible system details.
- An introduction of a Physics-Informed NeuDyE learning (PI-NeuDyE) to actively control the closed-loop accuracy of NeuDyE.
- Driving Port NeuDyE (DP-NeuDyE), a practical application of NeuDyE, reducing the number of inputs required for training.

Extensive case studies on the NPCC system validate the generalizability and accuracy of both PI-NeuDyE and DP-NeuDyE. These exhaustive analyses span a multitude of scenarios, differing in the time required for fault clearance, the specific fault locations, and the limitations imposed by the accessibility of only a small subset of data. Test results have demonstrated the scalability and practicality of NeuDyE, showing its potential to be used in ISO and utility control centers for online transient stability analysis and for planning purposes.

## 1. INTRODUCTION

Reliable discovery of dynamic equivalent models for unidentified subsystems, specifically external systems, is crucial to ensure reliable operations of large-scale interconnected transmission systems [1]. However, this task has been a longstanding challenge due to the existence of nonlinear dynamics, complex coherency characteristics, and unavailable component models in power systems [2][3]. Recent advancements in Phasor Measurement Units (PMUs) provide an opportunity to readily obtain rich history of high-data-rate measurements, which fostered the development of data-driven dynamic equivalence [4]. While the utilization of neural networks into power systems is not a new concept [5], [6], their application in power system dynamic equivalence remains an evolving area of research [7][8]. Early investigations primarily concentrated on the direct application of neural networks in power system analysis [9][10], or to use deep learning methods to unveil power system dynamics. Distinguished from other methods, Physics-Informed Neural Networks (PINNs) are engineered to directly leverage physical knowledge to assist the training procedure [11].

Despite various attempts being reported in the literature, significant challenges persist. First, learning continuous-time dynamic behaviors using discrete-time measurements poses a considerable obstacle. Traditional discretization techniques may not fully capture the intricacies of the continuous dynamics, leading to large inaccuracies that limit its practical implementations. Second, achieving robust and stable closed-loop operations under diverse operating conditions and disturbances is essential for safe plug-and-play integration of dynamic equivalence. Finally, minimizing the required measurements to ensure a feasible and practical implementation poses a challenge.

This research makes three significant contributions to address the aforementioned challenges:
- Formulation of Ordinary Differential Equations (ODEs)-Net-enabled Dynamic Equivalence (NeuDyE): This approach leverages ODEs and neural networks to model the system dynamics accurately, providing a continuous-time, data-driven representation that aligns with the actual behavior of power grids.
- Introduction of Physics-Informed Neural Dynamic Equivalence (PI-NeuDyE): It combines an ODE-NET-enabled equivalent model with physics-informed learning to identify a continuous-time dynamic equivalence while ensuring a close match in the closed-loop dynamic behaviors under disturbances.
- Implementation of a Driving Port NeuDyE (DP-NeuDyE): It reduces the number of inputs required for training, making it more manageable and cost-effective to deploy in real-world interconnected bulk power systems. Its generalization ability is also explored based on electrical distance.

This article is organized as follows: Section 2 introduces the mathematical basis of NeuDyE, i.e., how to formulate and simulate a power system with subsystems modeled by neural dynamic equivalence. Section 3 explains the key technology of PI-NeuDyE, which is to use Neural-Ordinary-Differential-Equation-Network (ODE-NET) to discover the dynamic equivalence of power systems. Section 4 introduces a variant of NeuDyE called DP-NeuDyE to further trim the features needed. Section 5 presents extensive case studies on the 140-bus NPCC system [12]. The results demonstrate the effectiveness of NeuDyE and its capability to handle various contingencies. Section 6 discusses the trade-off between the two methods and states the conclusion.

## 2. PROBLEM FORMULATION

For a reliability coordinator (RC), the entire interconnection can be partitioned into an internal system (InSys) and an external system (ExSys). An RC usually has both accurate dynamic models and real-time observability for InSys, but not ExSys. In reality, there may be a cushion area where the RC has partial observability; however, in this paper, we consider it part of the ExSys as well. Take the 140-bus NPCC system as an example, InSys and ExSys, connected through two tie lines [13], are illustrated in Figure 1.

InSys (bus1-36), which is the simplified ISO New England (ISO-NE) system, represents the subsystem that can be characterized by precise knowledge of its structure and parameters, enabling straightforward formulation using dynamic models for each component. The model-based InSys can be formulated based on the known dynamics of each component (e.g., the generators, loads, converter, transmission lines) by a set of differential-algebraic equations (DAE) in Equations (2.1)-(2.2).

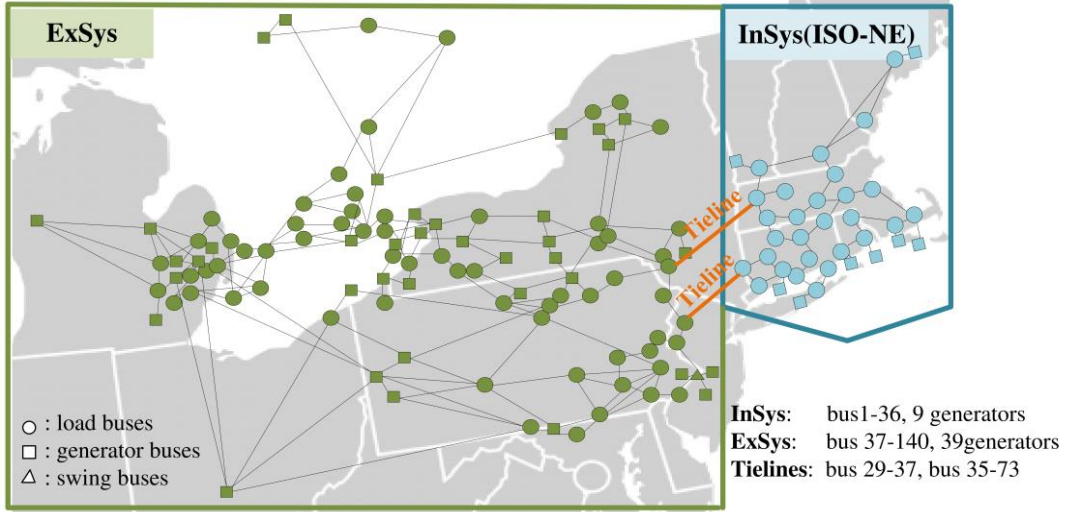

Figure 1 Topology of the NPCC system

In contrast, ExSys (bus37-140) lacks accessible physics models due to factors such as unavailable system state measurements, privacy concerns and inaccessible local measurements, e.g., real-time dispatch of the generators. Therefore, a data-driven neural network based dynamic equivalence is relied upon to model ExSys, as in Equation (2.3) [14]:

$$\frac{dx_{in}}{dt} = \mathcal{P}(x_{in}, y_{in}, i_{tie}) \tag{2.1}$$

$$G(x_{in}, y_{in}, i_{tie}) = 0 \tag{2.2}$$

$$\frac{dx_{ex}}{dt} = \mathcal{N}(x_{ex}, z_{in}) \tag{2.3}$$

Here, $x_{in}$ denotes the state variables of InSys's components (e.g., generators, turbines, exciters); $y_{in}$ denotes the algebraic variables of InSys such as power flow states; $i_{tie}$ denotes the currents flowing through the tie lines. Functions $\mathcal{P}$ and $G$ denote the dynamic and algebraic equations of InSys, respectively, which can be readily established based on the physics models of InSys. $x_{ex}$ denotes the state variables of ExSys; $z_{in}$ denotes the features from InSys, which is selected from part of the states of InSys to describe the interaction between InSys' dynamics and ExSys' dynamics. $\mathcal{N}$ is the forward propagation function of a neural network, which mimics the ExSys dynamics. This neural network is multi-layer structured, whose forward propagation can be functionally expressed as:

$$\mathcal{N}(x_{ex}, z_{in}) = \mathcal{L}_m(\mathcal{L}_{m-1}(\cdots \mathcal{L}_1(x_{ex}, z_{in}, \theta_1) \cdots, \theta_{m-1}), \theta_m) \tag{2.4}$$

where $\mathcal{L}_m$ denotes the loss function of the $m^{th}$ layer and $\theta_m$ denotes the corresponding weights in that layer. The universal approximation theorem ensures that a Deep Neural Network (DNN) can approximate any continuous functions for inputs within a specific range. Therefore, the advantage of a neural network-enabled dynamic equivalence lies in its flexibility for approximating a dynamic system without requiring the system to be linear or assuming any dynamical modes beforehand.

## 3. ODE-NET-ENABLED DYNAMIC EQUIVALANCE
### 3.1 NECESSITY OF CONTINUOUS-TIME LEARNING

A neural network can be regarded as a nonlinear function characterized by parameters $\theta$. In machine learning, these parameters are optimized by minimizing a loss function, typically computed as the error between measurements and neural network outputs. However, in this problem, the difficulty lies in the fact that the output of the neural network is the derivative of $x_{ex}$ while the measurements only provide $x_{ex}$, making the loss function hard to construct directly. To solve the problem, two different paths exist: discrete-time learning and continuous-time learning.

Conventional machine learning techniques for dynamic equivalence primarily rely on discrete-time learning. In discrete-time learning, the loss function for neural network training is usually constructed by discretizing the continuous-time differential equations into discrete-time difference equations. For example, based on the trapezoidal rule, the ExSys dynamics can be discretized as:

$$\frac{x_{ex}(t) - x_{ex}(t-\Delta)}{\Delta} = \frac{1}{2}(\mathcal{N}(x_{ex}(t), z_{in}(t)) + \mathcal{N}(x_{ex}(t-\Delta), z_{in}(t-\Delta))) \tag{3.1}$$

Correspondingly, the loss function can be established, and the neural network can be optimized by:

$$min_\theta \sum_{i=0}^{n} L_{discrete} = \sum_{i=0}^{n} \frac{1}{2}\eta_i \| \hat{y}_i - y_i \|_2$$
$$s.t. \quad \hat{y} = \frac{1}{\Delta}(\hat{x}_{ex}(t) - \hat{x}_{ex}(t-\Delta)), \quad y = \frac{1}{2}(\mathcal{N}(x_{ex}(t), z_{in}(t)) + \mathcal{N}(x_{ex}(t-\Delta), z_{in}(t-\Delta))) \tag{3.2}$$

where the subscript $i$ denotes the time-step, $n$ is the number of total time steps, $\hat{y}$ denotes the derivatives estimated from the measurements; and $y$ denotes the derivatives estimated from the neural network; $\eta_i$ denotes the weighting factor at time step $i$. However, this approach is sensitive to derivative estimation, resulting in biased training outcomes due to residue errors during training or non-ideal measurements. Although discrete-time training may produce satisfactory derivatives fitting, it cannot guarantee the accuracy of system states after numerical integration, leading to unsatisfactory performance in learning continuous-time dynamics.

Our solution is an ODE-NET-enabled dynamic equivalence, which adopts a continuous-time learning philosophy by directly minimizing the error between the state measurements $\hat{x}$ and the numerical solution of Equation (2.4):

$$min_\theta \sum_{i=0}^{n} L(x_{ex,i}) = \sum_{i=0}^{n} \frac{1}{2}\eta_i \| \hat{x}_{ex,i} - x_{ex,i} \|_2, \quad s.t. \quad \frac{dx_{ex,i}}{dt} = \mathcal{N}(x_{ex}, \hat{z}_{in}, \theta) \tag{3.3}$$

Comparing Equation (3.2) with (3.3), an obvious distinction is that ODE-NET is capable of directly minimizing the difference between real dynamic states and trained dynamic states, which requires no discretization and fully respects the continuous-time characteristics of power system dynamics. Therefore, it is theoretically less vulnerable to non-ideal measurements and the residue training error.

### 3.2 PHYSICS-INFORMED CONTINOUS-BACKPROGATION

Traditional DNNs are generally trained by the backpropagation technique, which computes the gradient of the loss function with respect to the DNN parameters at each layer to update the DNN. However, the ODE-NET training shown in Equation (3.3) differentiates from the conventional loss function, such as Equation (3.2) because it involves the numerical integration in its constraints.

To deal with the integration-incorporated constraints, ODE-NET adopts a continuous backpropagation technique to perform the neural network training. An adjoint method [14][8] is introduced to transform Equation (3.3) into a format that removes the numerical integration constraints; then a physic-informed (PI) continuous-backpropagation technique is developed as follows:

$$\min_\theta \sum_{i=0}^n L(x_{ex,i}, x_{in,i}) = \sum_{i=0}^n \frac{\eta_i}{2}(\|x_{ex,i} - x_{ex,i}\|_2 + \|x_{in,i} - x_{in,i}\|_2)$$

$$\mathcal{L} = \sum_{i=0}^n L_i - \int_{t_0}^{t_n}\left[\lambda^T(\dot{x}_{ex} - \mathcal{N}_\theta) + \mu^T(\dot{x}_{in} - \tilde{\mathcal{P}})\right]dt \qquad (3.4)$$

$$s.t.\ x_{ex,i} = x_{ex,0} + \int_{t_0}^{t_i}\mathcal{N}(x_{ex}, z_{in}, \theta)dt,\ x_{in,i} = x_{in,0} + \int_{t_0}^{t_i}\mathcal{P}(x_{in}, y_{in}, i_{tie})dt$$

Where $\lambda$ and $\mu$ respectively denote the adjoint states for ExSys and InSys, $\tilde{\mathcal{P}}$ is equivalently formulated from $\mathcal{P}$ using Equation (2.2). A dynamic equivalence is theoretically non-autonomous, where the InSys' states also impact the ExSys' dynamics. Thus, the dynamics of InSys and ExSys are both considered in Equation (3.4), which assures the performance of ODE-NET in the closed-loop simulation of the whole power system. With proper adjoint boundary conditions [8], the physics-informed gradient is:

$$\frac{d}{dt}\begin{bmatrix}\lambda^T \\ \mu^T \\ \partial\mathcal{L}/\partial\theta\end{bmatrix} = \begin{bmatrix}-\lambda^T\,\partial\mathcal{N}/\partial x_{ex} - \mu^T\,\partial\tilde{\mathcal{P}}/\partial x_{ex} \\ -\lambda^T\,\partial\mathcal{N}/\partial x_{in} - \mu^T\,\partial\tilde{\mathcal{P}}/\partial x_{in} \\ \lambda^T\,\partial\mathcal{N}/\partial\theta\end{bmatrix} \qquad (3.5)$$

$$\theta^* \leftarrow \theta - \frac{\partial\mathcal{L}}{\partial\theta}$$

Finally, the gradient descent for $\mathcal{N}_\theta$ can be performed using $\partial\mathcal{L}/\partial\theta|_{t=0}$ integrated from Equation (3.5) by any ODE solvers. As illustrated in Figure 2(a), PI-NeuDyE explicitly embeds the accuracy of both ExSys and InSys states in a closed-loop manner, which ensures NeuDyE generates dependable dynamic responses in conformance with the system's real dynamics once the training converges.

## 4. SEEN FROM DRIVING PORT EQUIVALENCE

In the previously devised PI-NeuDyE, the optimal closed-loop results are achieved using extensive inputs from the internal system, which may not be readily available in practical applications [16]. Another notable concern is that the PI-NeuDyE, while considering the interaction between the ExSys and InSys in a closed-loop fashion, ensures accuracy at the expense of compromised efficiency. To address these limitations and enhance the method's applicability in real-world scenarios, we introduce a more scalable neural equivalent technique known as the Driving Port (DP) NeuDyE. This novel approach necessitates only the knowledge of boundary voltages for the tie-lines, thereby facilitating the practical implementation of NeuDyE in utility and industrial systems.

### 4.1 ALGEBRAIC COMPONENT SEPARATION

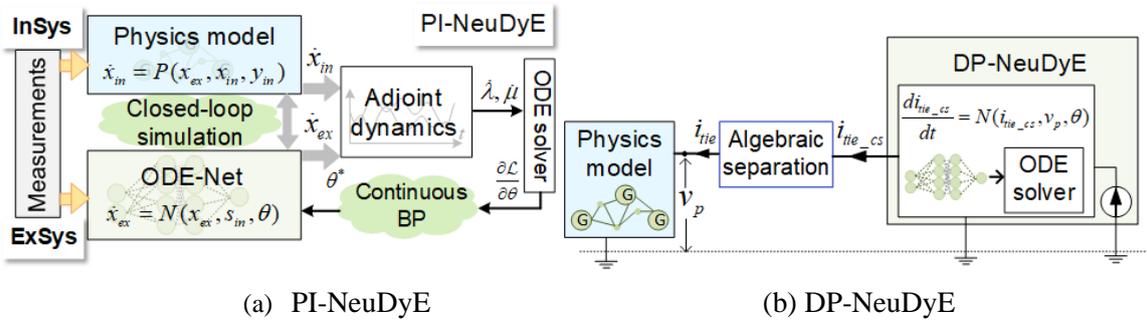

(a) PI-NeuDyE  (b) DP-NeuDyE

Figure 2 Network equivalent methodologies

To form the neural network-integrated power grid, following Equations (2.1)-(2.3), the selection of $x_{ex}$ and $z_{in}$ is important. If ExSys is static, a Norton equivalent current source, depicted in Figure 2(b), can replace it. From the perspective of InSys, the representation of ExSys in full detail or as a Norton equivalent current source yields the same output $i_{tie}$ for the given input $v_p$. Inspired by the Norton equivalent theory, a neural network observed from the driving port is devised.

To capture its nonlinear dynamics, measurements of port voltages $v_p$ and tie currents $i_{tie}$ are utilized to discover the state space model of ExSys as shown in Figure 2(b). The tie currents $i_{tie}$ are selected as the state variables for external system $x_{ex}$, whose continuous differential structure is represented by the neural network; then an algebraic component separation, is introduced bellow:

$$\frac{di_{tie}}{dt} = \mathcal{N}(i_{tie}, v_p, \theta)$$
$$\mathcal{G}(i_{tie}, v_p) = 0 \tag{4.1}$$

where the port voltages $v_p$ corresponds to the InSys features $z_{in}$ in Equation (2.3). The tie currents $i_{tie}$ can then be represented as a linear combination of state variables and inputs:

$$i_{tie} = C_s \cdot x_{ex} + D \cdot v_p \tag{4.2}$$

where matrices $C_s$ and $D$ are constant matrices. If a fault happens in the internal network at time instant $t_i$, sudden changes may happen in $\Delta v_p(t_i) = v_p(t_i + \Delta t) - v_p(t_i)$, while $x_{ex}(t_i)$ keeps invariant in a very short period of $\Delta t$, i.e. $\Delta x_{ex}(t_i) = 0$. The components $i_{tie}$ can be split into two types of components: continuous-state-variable components $i_{tie\_cs} = C_s \cdot x_{ex}$ and algebraic components $i_{tie\_a} = D \cdot v_p$. On one hand, algebraic components embody the immediate contribution from the port voltages $v_p$, which may exhibit discontinuity during switching events within the internal network. On the other hand, continuous-state-variable components, $i_{tie\_cs}$, are employed as the constituents of the neural network equivalent, as illustrated in Equation (4.1). These components fulfill the need for continuity in Equation (4.1).

To compute the coefficient matrix $D$, we leverage measurement data obtained during the fault period. This is achieved by using the least squares method, as shown below:

$$\begin{bmatrix} D_{11} & \cdots & D_{1n_p} \\ \vdots & \ddots & \vdots \\ D_{n_p 1} & \cdots & D_{n_p n_p} \end{bmatrix} = \begin{bmatrix} \Delta i_{tie}(t_1), & \cdots, & \Delta i_{tie}(t_{n_f}) \end{bmatrix} \cdot \begin{bmatrix} \Delta v_p(t_1), & \cdots, & \Delta v_p(t_{n_f}) \end{bmatrix}^{-1} \tag{4.3}$$

where $n_p$ is the number of port voltages and $n_f$ is the number of faults whose port voltages and tie line currents are recorded in the data sets. Define $\Delta v_p(t_i) = v_p(t_i + \Delta t) - v_p(t_i)$ and $\Delta i_{tie}(t_i) = i_{tie}(t_i + \Delta t) - i_{tie}(t_i)$. The continuous component $i_{tie\_cs}$ can then be extracted as:

$$i_{tie\_cs} = i_{tie} - D \cdot v_p \tag{4.4}$$

The neural equivalent network in Equation (4.1) and the DAE now become:

$$\frac{di_{tie\_cs}}{dt} = \mathcal{N}(i_{tie\_cs}, v_p, \theta)$$
$$\frac{dx_{in}}{dt} = \mathcal{P}(x_{in}, y_{in}, i_{tie})$$
$$\mathcal{G}(x_{in}, y_{in}, i_{tie\_cs}, i_{tie}, v_p) = 0 \tag{4.5}$$

The neural equivalent of the external system and the corresponding formulated interface as shown in Figure 2(b) are integrated into Transient Stability Analysis (TSA) simulation. The equivalent admittance is formed by the coefficient matrix $D$ in Equation (4.4). The values of current sources are updated by applying an explicit integration method to Equation (4.5).

## 4.2 FORMULATION OF ODE-NET BASED DRIVING PORT EQUIVALENCE

In real-world power networks, measurement data is inherently discrete, even though the underlying system is continuous. Therefore, DP-NeuDyE is again discovered through a continuous-time learning manner, as depicted in Equation (4.5), from discrete measurements $i_{tie}$ and $v_p$. During the time interval from $[0, t_n]$, DP-NeuDyE is trained by minimizing the loss function defined by the error between the state measurements $\hat{i}_{tie\_cs}$ and the numerical solution $i_{tie\_cs}$ by Equation (4.5), as illustrated below:

$$L = \sum_{i=0}^{n} \frac{1}{2} \| i_{tie\_cs}(t_i) - \hat{i}_{tie\_cs}(t_i) \|_2$$

$$s.t.\ i_{tie\_cs}(t_i) = \hat{i}_{tie\_cs}(0) + \int_0^{t_i} \mathcal{N}(x, u, \theta) dt$$

(4.6)

Similar as in Equation (3.5), the challenge in minimizing Equation (4.6) stems from the integration operation in the constraints. ODE-NET tackles this challenge by treating the ODE solver (the integration operation) as a black box and computing gradients using the adjoint sensitivity method [15][16].

## 4.3 STRENGTHENING DP-NEUDYE VIA RECURRENT NEURAL NETWORK

Data-driven methodologies predominantly depend on observable states to construct the neural equivalent model, as exemplified in Equation (4.1). However, this model reduction approach inherently leads to a scenario where state variables constitute only a minor subset of the comprehensive state variables present in the original power network. As a result, such a reduction may not entirely encapsulate all the crucial dynamic properties intrinsic to the power system. To address this deficiency of information, DP-NeuDyE is enhanced by leveraging historical data through the implementation of Recurrent Neural Networks (RNNs).

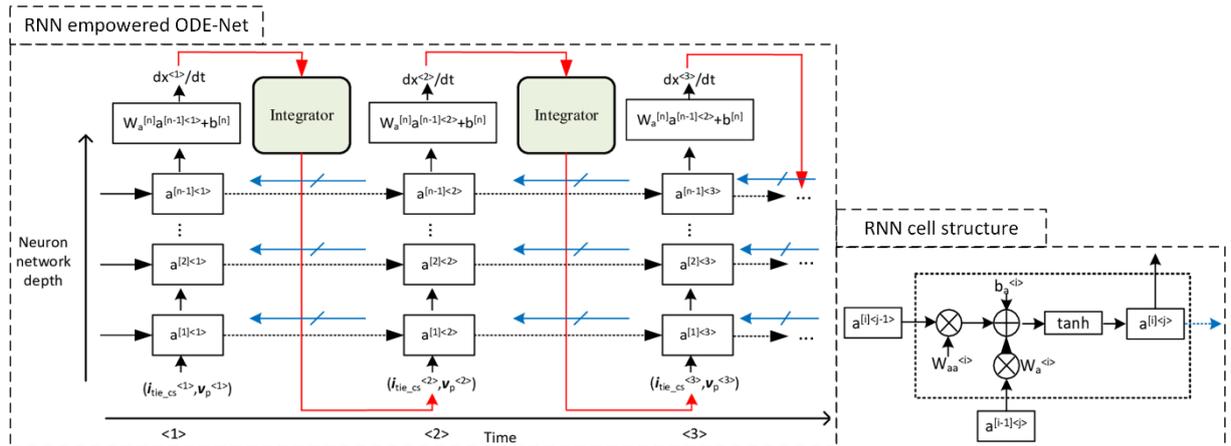

Figure 3 RNN empowered ODE-NET

RNNs, with their unique ability to remember past information, provide a robust mechanism to incorporate temporal dynamic behavior into the model [17]. This allows for a more comprehensive understanding of the system dynamics, thereby improving the model's performance. The integration of RNNs into the DP-NeuDyE framework is illustrated in Figure 3 and explained as follows. In Figure 3, the structure of the RNN cell integrates the output of a specific neuron from the previous time step into the computation of the current time step's output for the same neuron. This mechanism effectively leverages historical data from the preceding time step to assist in calculating the derivative of the current time step. Therefore, this approach ameliorates potential information deficiencies that might arise when computations rely on a limited subset of state variable, thereby bolstering the overall accuracy and robustness of DP-NeuDyE.

Similar in Section 3, the backward propagation should be used to evaluate the gradient of the neural network parameters. Recall that the continuous backpropagation for ODE-NET in Subsection 3.2, already considers integration along the time by solving an augmented differential equation. Therefore, the backward propagation throughout time for the RNN cell is ignored and the gradient descent method used in Subsection 3.2 can directly be applied to the RNN-empowered ODE-NET.

## 5. CASE STUDY

In this section, the detailed training and testing procedures of NeuDyE are introduced. Simulation results of PI-NeuDyE and DP-NeuDyE are presented to demonstrate their efficacy and practicality.

### 5.1 ALGORITHM SETTINGS

The ground truth electromechanical trajectories are obtained by simulating the complete, physics-based 140-bus NPCC system via the Power System Toolbox (PST) [18]. The PST results are verified with simulations from Transient Security Assessment Tool (TSAT). Trapezoidal rule is adopted as the numerical integration method for the devised method.

#### 5.1.1 NeuDyE Training

PI-NeuDyE: as introduced in Section 3, the selected states are from generators, exciters, governors, and line currents of InSys as $s_{in}$, in total 90 dimensions; the tie line currents are the states of ExSys $x_{ex}$ with 4 dimensions (2 tie lines, each has a real part and an imaginary part). Note that such training features can be flexibly adjusted according to available measurements. If the training features are abundant, the training process might be slower because of the numerical integration burden.

DP-NeuDyE: followed by Section 4, $s_{in}$ includes boundary voltages with 4 dimensions (2 ports, each with a real part and an imaginary part); $x_{ex}$ consists of tie line currents; as such, DP-NeuDyE is faster than PI-NeuDyE.

#### 5.1.2 NeuDyE Testing

Once an ODE-NET is acquired, its performance is evaluated through closed-loop tests. In the closed-loop context, the ODE-NET-based ExSys model takes the place of the large and unknown external systems. It integrates with the physics-based InSys model, forming a physics-neural-integrated system that operates as a unified whole. This constitutes the final setup, wherein the dynamics of the entire system are computed through numerical integration. The predicted values are trajectories simulated by the physics-neural-integrated system, which contains the 36-bus, physics based InSys and the ODE-NET-based dynamic equivalence of the ExSys. If the NeuDyE can accurately mimic the dynamics of the original ExSys, the predicted values should be close to the true values, i.e., simulation with the full system model.

### 5.2 SIMULATION RESULTS

#### 5.2.1 Validity of PI-NeuDyE under various fault clearing times and fault locations

As mentioned in Section 3, traditional discrete learning may yield satisfactory predictions in open-loop training by fitting the derivatives, like in Figure 4(a), which is the prediction from a deep neural network (DNN). Whereas in the closed-loop test in Figure 4(b), DNN fails to capture the continuous dynamics after the integration. The biased training emerges from accumulated residue errors during the training process. Even when the discrete-learning DNN yields flawless predictions of derivatives in open-loop training, the integrated results in the closed-loop test turn out to be impractical. This underscores the imperative for continuous-time learning.

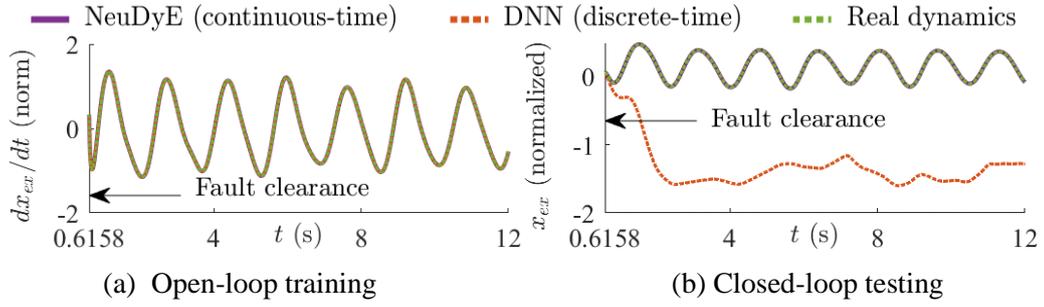

(a) Open-loop training  (b) Closed-loop testing

Figure 4 Comparison of NeuDyE with conventional discrete-time DNN

Depicted in Figure 5, 25 training scenarios are generated by launching three-phase faults at 0.50s at bus 18, 19, 20, 21, or 28 with fault clearing set randomly within a time interval [0.53s, 0.6s]. The training variables of InSys have 90 dimensions as mentioned in Subsection 5.1.1

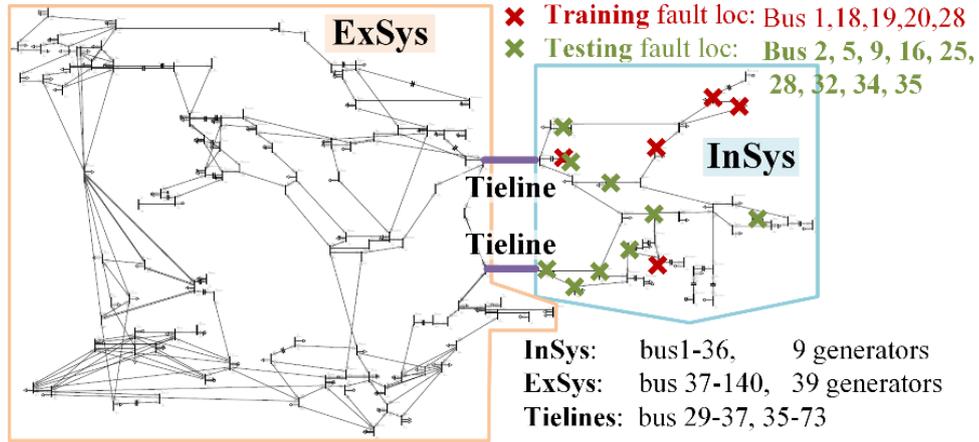

Figure 5 140-bus NPCC system training and testing locations.

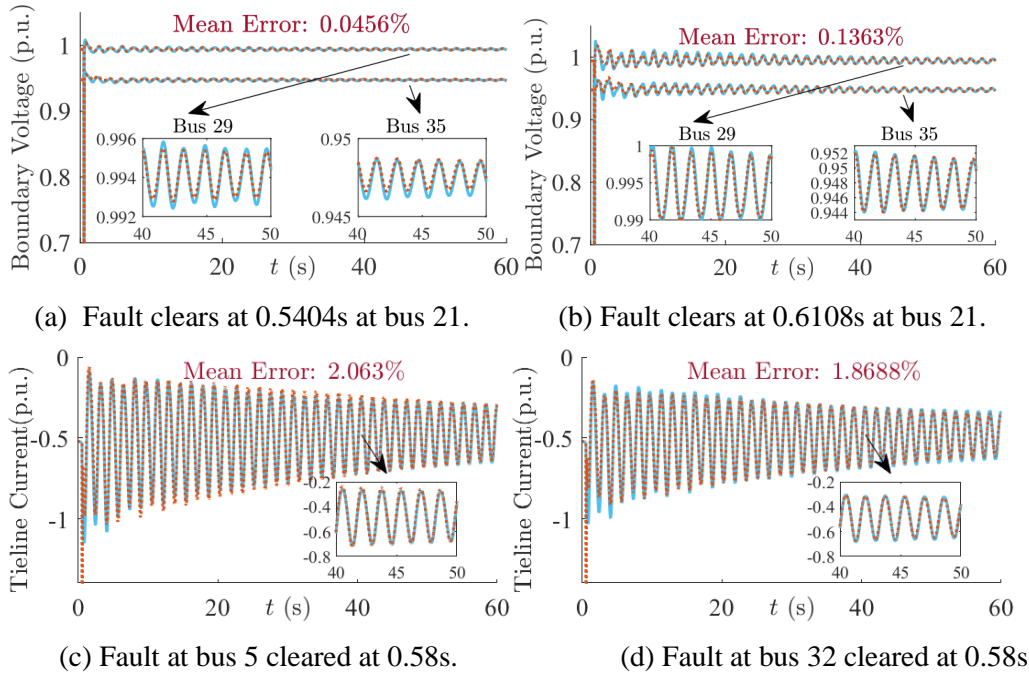

(a) Fault clears at 0.5404s at bus 21.  (b) Fault clears at 0.6108s at bus 21.

(c) Fault at bus 5 cleared at 0.58s.  (d) Fault at bus 32 cleared at 0.58s

Figure 6 Closed-loop test results of fault on different locations with different clearing times

Figure 6(a)-(b) show the closed-loop test results of boundary voltages with faults on bus 21 cleared at 0.5405s, 0.6108s, which are new values to the training sets. The perfect match illustrates the accuracy of the developed method under varied fault clear times. In Figure 6(c)-(d), test results of tieline currents with faults on bus 5 and bus 32 demonstrate a perfect match between PI-NeuDyE's results and that from the full NPCC model.

Further, in Figure 7, 108 testing scenarios are generated with new fault locations and random fault clearing times at bus 2, 5, 9, 16, 25, 28, 32, 34 and 35. The box plot shows that the overall relative error is lower than 1%, indicating a satisfying generalization ability. Figures 6-7 show that the derived PI-NeuDyE model can properly and accurately represent the dynamics and transients, regardless of changes in fault durations or fault locations.

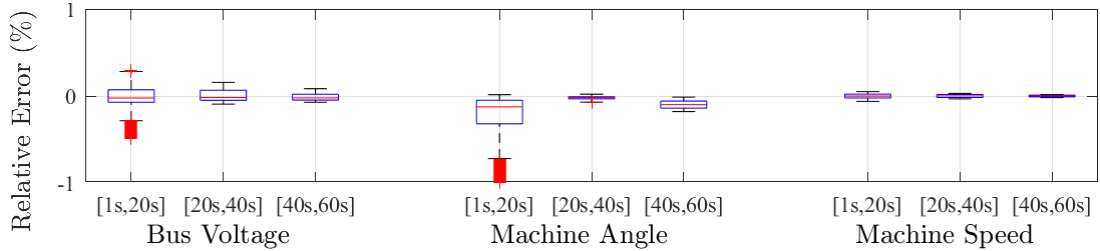

Figure 7 Accuracy of PI-NeuDyE under 108 cases with new fault locations and random clearing times

As mentioned in Subsection 5.1, choosing features of InSys is flexible. Other settings, like using active and reactive power of transmission lines or bus voltages and line currents, are also feasible. As an example, Figure 8(a) presents a new testing result by using boundary voltages between ExSys and InSys and all branch currents in InSys as training features (90 dimensions for InSys). Fig. 8(b) uses boundary currents and voltage from all buses (76 dimensions for InSys). The fault clears at 0.56s at bus 2. The predicted trajectories match with the measurements closely, with a mean error of less than 0.4%, indicating high accuracy and satisfactory performance of NeuDyE using different settings of training features.

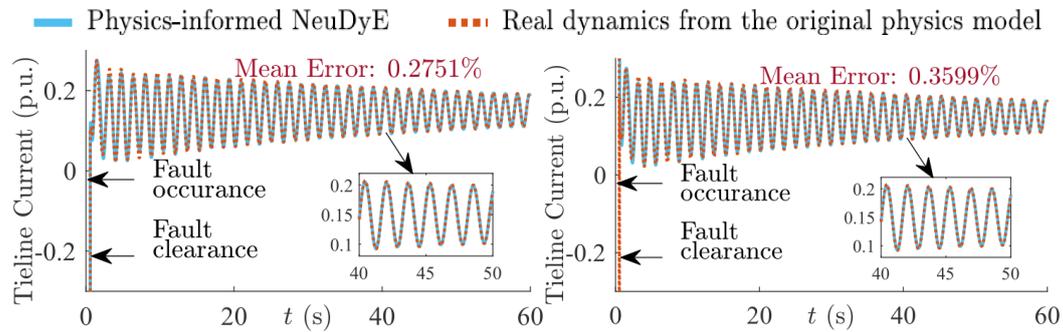

(a) Features: all branch currents and boundary voltages. (b) Features: boundary currents and all voltages.

Figure 8 Performance of NeuDyE using other measurements.

### 5.2.2 Reduced variables using DP-NeuDyE

As previously mentioned, DP-NeuDyE is designed for potential practical applications that requires limiting the number of input variables. In contrast with Subsection 4.2, where the number of InSys features used in Figures 6-7 are 90 dimensions, DP-NeuDyE only needs 4 dimensions of InSys features. The selections of ExSys features are the same for both methods. The ExSys subsystem, as depicted in Figure 1, is modeled by the DP-NeuDyE as illustrated in Figure 2(b).

The training trajectories are derived from faults happening in five distinct buses, each triggered by phase-to-ground faults at the T-line, as highlighted in red in Figure 9. All the faults occur at 0.5s and clear at 0.55s.

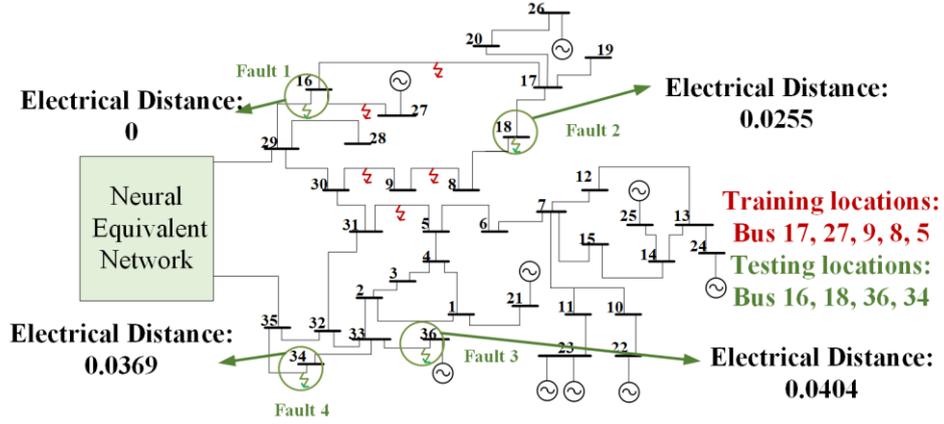

Figure 9: Training and testing locations for DP-NeuDyE

The testing locations are shown in green. In Figure 10, it is evident that DP-NeuDyE adeptly predicts low-frequency oscillations across a considerable domain. The ability to precisely forecast scenarios beyond the training set serves as validation for the efficacy of the proposed method. Furthermore, the full model-based trajectory exhibits two oscillation modes: one at 0.5991 Hz with a magnitude of 0.2155 and another at 1.24813 Hz with a magnitude of 0.1349. The simulation based on the neural equivalent model also accurately predicts these oscillation modes, with magnitudes of 0.2051 and 0.1431, respectively.

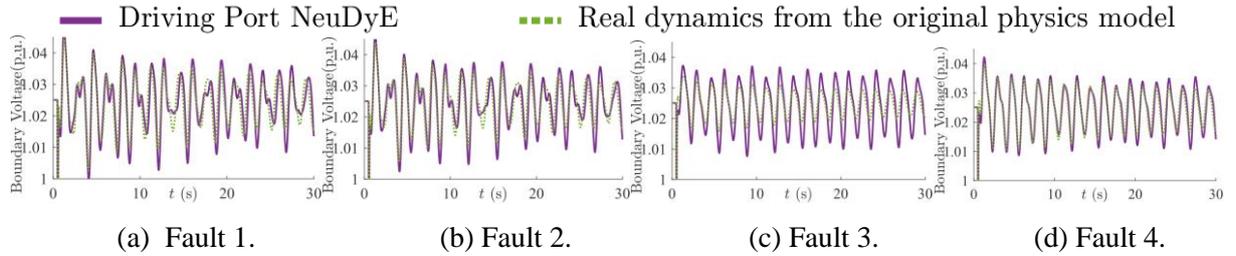

(a) Fault 1.    (b) Fault 2.    (c) Fault 3.    (d) Fault 4.

Figure 10 DP-NeuDyE closed-loop simulation.

### 5.2.3 Generalizability analysis based on electrical distance

To quantify the NeuDyE models' generalization performance, we employ the electrical distance between the fault locations in the testing set and those in the training set as a measure. The network topology is transformed into an adjacency matrix using graph theory, as depicted in Equation (5.1), thereby establishing an automated method for predicting the performance of the introduced neural ODE model for subsystems. Consequently, the electrical distance between a new fault location and those in the training set can be determined from the adjacency matrix by selecting the shortest distance.

$$\text{Ajacency matrix } A : \begin{cases} \text{Electric connection between BUS i and j: } A_{ij} = X_{ij} \\ \text{Else: } A_{ij} = 0 \\ \text{Fault dynamic record near BUS i: } A_{ii} = 1 \end{cases} \quad (5.1)$$

where $i \neq j$ and $X_{ij}$ is the reactance (p.u.) in the branch connecting $BUS\ i$ and $j$.

In the previous case study as depicted in Figure 10, the electrical distances from the test set to the training set are as follows: 0, 0.0255, 0.0404, 0.0369. These electrical distances are relatively small; the generalizability of the DP-NeuDyE is thus relatively good. In terms of the performance in predicting boundary voltage, Figure 10(c) is already not satisfiable compared with results from PI-NeuDyE in Figure 6. If a fault occurs at a considerably remote distance from those in the training set, such as the T-line between bus 19 to 17 in Figure 11, where the electrical distance is 0.0948, DP-NeuDyE may output divergent dynamic predictions. Even though RNN-empowered DP-NeuDyE yields convergent results, it still faces challenges in achieving precision.

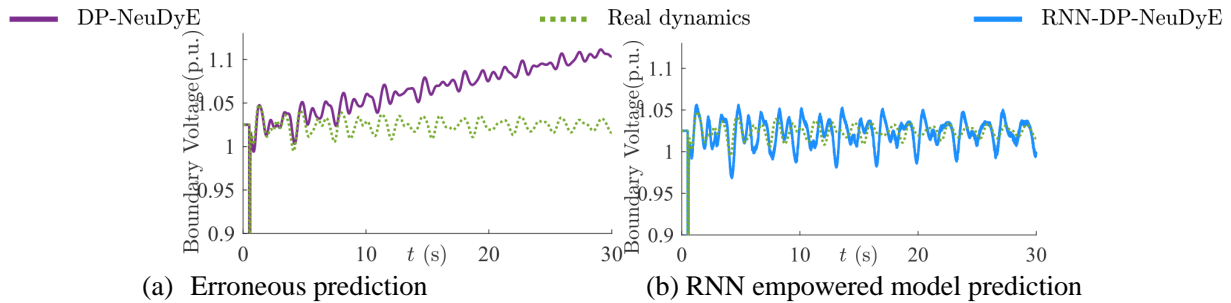

(a) Erroneous prediction  (b) RNN empowered model prediction

Figure 11: Testing results of DP-NeuDyE under distant fault.

## 6. DISCUSSION

Table I: Training time for different methods

| Method | Training time/ iteration | Total iters till convergence |
| --- | --- | --- |
| PI-NeuDyE | 49.1706s | 3127 |
| DP-NeuDyE | 20.4417s | 3322 |
| RNN-DP-NeuDyE | 103.2257s | 2772 |

From Table I, for the same training set, it is obvious that DP-NeuDyE is more efficient. The primary advantages in application lie in DP-NeuDyE's ability to perform dynamic equivalencing even when some features in InSys are not available. It is noteworthy that DP-NeuDyE utilizes only 6% of the inputs compared to PI-NeuDyE. This capability eliminates the requirement for extensive data acquisition and storage resources. However, the increased training time for RNN empowered DP-NeuDyE is notable. This is attributed to the fact that it employs RNN-ODE-Net for training, while PI-NeuDyE and DP-NeuDyE employ ODE-Net with a Multilayer Perceptron (MLP) structure. The inherently slower training speed of RNN compared to MLP, even with the same network size, contributes to the observed increase in training time.

Another observation is the performance of DP-NeuDyE is not as commendable as PI-NeuDyE for faults occurring from a large electrical distance. The primary reason behind this discrepancy lies in the training process. PI-NeuDyE trains in a closed-loop manner by considering the interacting dynamics of both InSys and ExSys, involving 90 dimensions of InSys features. Whereas DP-NeuDyE only sees from the driving port, utilizing only 4 dimensions of boundary measurements as InSys features. As a result, there exists a trade-off between training efficiency and generalization ability, which impacts the overall performance of the DP-NeuDyE. Even though the utilization of RNN can improve the generalization capability of DP-NeuDyE, further investigation is required to address its precision and the time-consuming nature of its training.

To summarize, for faults not too distant from the training sets, both DP-NeuDyE and PI-NeuDyE yield satisfactory results. In cases where a fault is significantly distant from the training set, PI-NeuDyE remains effective, while DP-NeuDyE proves inadequate unless the training set is expanded across a wider area.

## 7. CONCLUSION

This article introduces a physics-informed Neural Dynamic Equivalence (PI-NeuDyE) and its practical variant Driving Port NeuDyE (DP-NeuDyE), which uncovers a powerful continuous-time dynamic equivalence of external systems. The effectiveness of DP-NeuDyE and PI-NeuDyE is demonstrated through case studies conducted on the 140-bus NPCC system, showcasing their performance under various fault locations and clearing times. Furthermore, comparisons are made between DP-NeuDyE and PI-NeuDyE in terms of efficiency and generalization ability.

In the future, the authors aim to enhance method efficiency by incorporating advanced computing technologies, sparsity techniques for power grid formulation and time-domain simulation.